\documentclass[structabstract]{aa}  

\usepackage{enumerate}
\usepackage{graphicx}
\usepackage{txfonts}
\usepackage{natbib}

\newcommand{\boldnabla}{\mbox{\boldmath$\nabla$}}

\begin{document}
  
  \title{MHD dynamical relaxation of coronal magnetic fields}

  \subtitle{IV. 3D tilted nulls}

  \author{Jorge Fuentes-Fern\'andez and Clare E. Parnell}

  \institute{School of Mathematics and Statistics, University of St Andrews, North Haugh, St Andrews, Fife, KY16 9SS, Scotland}

  \date{}

  \abstract
  {There are various types of reconnection that may take place at 3D magnetic null points. Each different reconnection scenario must be associated with a particular type of current layer.} 
  {A range of current layers may form because the topology of 3D nulls permits currents to form by either twisting the field about the spine of the null or by folding the fan and spine into each other. Additionally, the initial geometry of the field can lead to variations in the currents that are accumulated. Here, we study current accumulations in so-called 3D `tilted' nulls formed by a folding of the spine and fan. A non-zero component of current parallel to the fan is required such that the null's fan plane and spine are not perpendicular. Our aims are to provide valid magnetohydrostatic equilibria and to describe the current accumulations in various cases involving finite plasma pressure. }
  {To create our equilibrium current structures we use a full, non-resistive, Magnetohydrodynamics (MHD) code so that no reconnection is allowed. A series of experiments are performed in which a perturbed 3D tilted null relaxes towards an equilibrium via real, viscous damping forces. Changes to the initial plasma pressure and to magnetic parameters are investigated systematically.}
  {An initially tilted fan is associated with a non-zero Lorentz force that drives the fan and spine to collapse towards each other, in a similar manner to the collapse of a 2D X-point. In the final equilibrium state for an initially radial null with only the current perpendicular to the spine, the current concentrates along the tilt axis of the fan and in a layer about the null point with a sharp peak at the null itself. The continued growth of this peak indicates that the system is in an asymptotic regime involving an infinite time singularity at the null. When the initial tilt disturbance (current perpendicular to the spine) is combined with a spiral-type disturbance (current parallel to the spine), the final current density concentrates in three regions: one on the fan along its tilt axis and two around the spine, above and below the fan. The increased area of current accumulation leads to a weakening of the singularity formed at the null. The 3D spine-fan collapse with generic current studied here provides the ideal setup for non-steady reconnection studies.}
  {}

  \maketitle


\section{Introduction}

Magnetic null points (locations where all three components of the magnetic field are zero) have been associated with sites of magnetic reconnection ever since the notion of magnetic reconnection was first conceived \citep{Parker57,Sweet58}. It is now known that, in three dimensions (3D), they are not the only possible location for reconnection \citep[e.g.,][]{Schindler88,Hesse88,Hornig03,Priest03}, which may also take place in magnetic separators \citep{Longcope96,Longcope01,Haynes07,Parnell08,Parnell10a,Parnell10b} and quasi-separatrix layers \citep{Priest95,Demoulin96,Demoulin97,Aulanier06,Restante09,Wilmot09}. However, 3D null points still remain key sites for current accumulation and energy release. Using SOHO MDI (Michelson Doppler Imager) magnetogram data, \citet{Longcope09} estimate that there are around 20,000 nulls above 1.5 Mm in the solar atmosphere during solar minimum. Moreover, due to the rapid fall off in complexity of the magnetic field above the photosphere, more nulls are likely to occur below 1.5 Mm than are above. 

Over the years 3D magnetic null points have been studied in detail on a number of different levels. The first set of papers considers the basic structure of nulls \citep[e.g.][]{Cowley73, Fukao73} with a comprehensive account of the linear structure of all possible 3D nulls provided in \citet{Parnell96}. The general geometry of 3D magnetic null points involves two important features. The field lines that extend into/out of the null itself lie either along a pair of lines, known as {\it spines}, or on a surface, known locally as the {\it fan}. The field lines in the fan are all directed either away from the null, forming what is known as a positive (or B-type) null, or into the null, creating a negative (A-type) null. In the case of a positive null the spines are directed into the null and for a negative null they are directed away.

A second set considers the nature of possible reconnection scenarios at 3D nulls driven by plasma flows of varying types, which may depend on the behaviour of the flow pattern relative to the spine and fan. These perturbations may be caused by a local driver \citep{Rickard96,Bulanov02,Pontin04,Pontin05a,Pontin07a,Pontin07b,Pontin07c,Priest09,Masson09,Wyper10,Alhachami10,Pontin11} or by an indirect perturbation \citep{Pariat09,Pariat10,Edmondson09,Masson09,Masson12} far away from the null. In particular, flow patterns that twist the field about the spine, known as torsional reconnection regimes, dissipate currents that lie parallel to the spine and result in the slippage of the field lines about the spine. Flow patterns across the fan or spine, known as fan-spine reconnection regimes, are associated with currents perpendicular to the spine and cause a shearing of the spine and/or fan leading to reconnection at the null itself. 

In general terms, reconnection studies consider potential 3D nulls that are driven by some external force initiating reconnection. However, an arguably more realistic scenario is that flows in the system generate currents resulting in the collapse of the null in a particular way, since non-potential nulls are generically unstable \citep{Parnell97}, and leading to the accumulation of current at, or in the vicinity of, the null. Papers that look at the accumulation of currents and their associated equilibria form a third set of papers \citep[e.g. ][]{Pontin05b,Fuentes12c}. This paper belongs to this set. 

Current accumulations due to spiral nulls in which the field twists about the spine were studied in \cite{Fuentes12c}, the previous paper to this one in the series on `Dynamical relaxation of coronal magnetic fields'. We looked at the formation of a static non-force-free equilibrium due to a torsional-type disturbance at a 3D null point. In that paper, current accumulations were found due to the viscous non-resistive MHD relaxation of a non-potential null with a spine-aligned current density. In the present paper, we use the same approach, but here we study current accumulations due to a shearing-type disturbance to a 3D null in which the current and spine fold towards on another. Here we consider a non-potential null with a fan-aligned current density. The current accumulations formed must be different to those found in \cite{Fuentes12c} since the reconnection in the two cases is completely different.

\citet{Pontin05b} has considered the case of a such a perturbation, associated with a component of the current that is strictly perpendicular to the spine. They find a collapse of the fan and spine towards each other and the formation of a current singularity at the location of the null in a non-force-free equilibrium. In their paper, they consider the nature of this singularity and find that it scales with the numerical resolution in an equivalent manner to the 2D singularities studied by \citet{Craig05}. Also, they find that increasing the pressure does not stop the formation of the singularity, but does significantly reduce its strength. Both of these studies used a frictional Lagrangian code with a fictitious damping term, $-\kappa{\bf v}$, added to the momentum equation to enable the field to relax to an equilibrium. This model imposes a natural constraint, since their damping term is not physical. Indeed, the polytropic model with $p\approx \rho^\gamma$ is used for pressure, which imposes a condition of adiabaticity to the relaxation process.

\citet{Fuentes11} considered very similar, but subtly different, experiments to \citet{Craig05} involving the collapse of non-potential 2D X-points embedded within a plasma. One key difference is that \citet{Fuentes11} were able to follow the dynamical evolution of the system and the energetics, by using an MHD relaxation driven by viscous forces, which naturally has an associated heating term in the energy equation. Thus, the system does not evolve adiabatically. While \citet{Craig05} focus on the scaling laws for the formation of the singularity, \citet{Fuentes11} focus on the actual non-force-free equilibrium obtained due to the presence of a plasma pressure and the nature of the singularity as it grows in time. \citet{Fuentes11} show that the field left after the viscous relaxation has finished is in a ``quasi-equilibrium'' state where the central current layer, which stretches out a short way along the separatrices, slowly changes its shape becoming thinner and shorter. Indeed, the system is found to have entered an asymptotic regime and is heading towards an infinite time singularity.

The approach of \citet{Fuentes11} is quite different to the frictional, adiabatic relaxation of \citet{Craig05}. Similarly, the study we carry out here uses a different relaxation to the frictional, adiabatic approach of \citet{Pontin05b} because it involves the viscous, non-resistive, non-adiabatic relaxation of 3D nulls. Furthermore, the model we consider here differs from \citet{Fuentes12c} because here the collapse of the null and current accumulation is associated with a current that is purely perpendicular to the spine as opposed to parallel to the spine. Additionally, here we also consider the MHD relaxation associated with a more general current that has components both perpendicular and parallel to the spine, unlike either of the previous papers. 

More specifically, the aim of this paper is to investigate the nature of current accumulations at 3D radial magnetic nulls following a collapse due to the presence of an initial homogeneous current density whose component perpendicular to the spine is non-zero. This may be thought of as the result of a shearing-fan disturbance. The characteristics of any infinite time singularities will be determined, if they arise. We evaluate the effects of the plasma pressure in the evolution, while both the initial disturbance and the background plasma pressure are changed systematically. This work is a continuation of the work carried out in \citet{Fuentes10,Fuentes11} and \citet{Fuentes12c} on non-resistive MHD relaxation of magnetic fields embedded in non-zero beta plasmas.

The paper is structured as follows. In Sec. \ref{sec3}, we present the equations that define the initial configuration and the details of the numerical experiments. The results for 3D tilted nulls that are initially radial and have a fan-aligned current are presented in Sec. \ref{sec4}, while, in Sec. \ref{sec5}, the more general results for a radial null with components of current that are both parallel and perpendicular to the spine can be found. Finally, we conclude with a general overview of the problem in Sec. \ref{sec7}.


\section{Initial magnetic field configurations and numerical scheme} \label{sec3}

The initial magnetic field is chosen to be that of a positive linear 3D null point located at the origin of a constatnt current and associated with it. Like the initial field in \citet{Fuentes12c}, the spine of the null is chosen to be along the $z$-axis, however unlike \citet{Fuentes12c} the initial magnetic field here has a component of current perpendicular to the spine (along the $x$-axis and parallel to the fan), rather than parallel to the spine. In the first set of experiments, the current is purely perpendicular to the spine, and in the second set it has a general current with components both parallel and perpendicular to the spine.
 
The current density is chosen to be constant and can be written in terms of the component parallel to the fan, $j_f$, and parallel to the spine, $j_{sp}$, as
 \begin{equation}
{\bf j}=\frac{1}{\mu}(j_f, 0, j_{sp})\;.\nonumber
\end{equation}
Following \citet{Parnell96} the magnetic field, ${\bf B}$, is
\begin{equation}
\label{M_general}
{\bf B} = \left(x-\frac{j_{sp}}{2}y,\frac{j_{sp}}{2}x+by,j_fy-(b+1)z\right)\;,
\end{equation}
where $b>0$ in order to ensure that the spine lies along the $z$-axis and that the null is positive.

\begin{figure*}[t]
  \begin{minipage}[b]{1.0\linewidth}

    \begin{minipage}[b]{0.50\linewidth}
      \centering
      \includegraphics[scale=0.32]{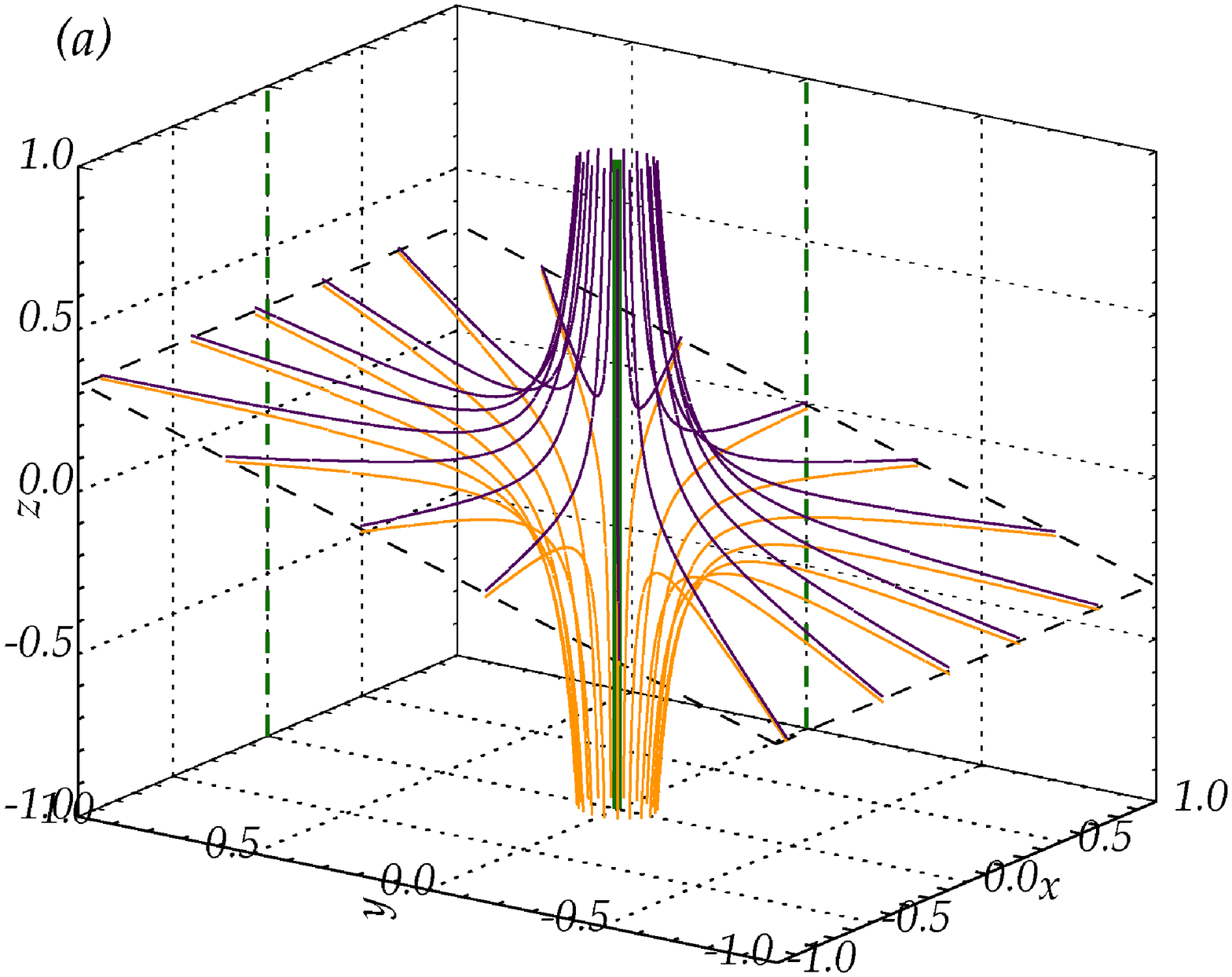}
    \end{minipage}
    \begin{minipage}[b]{0.50\linewidth}
      \centering
      \includegraphics[scale=0.30]{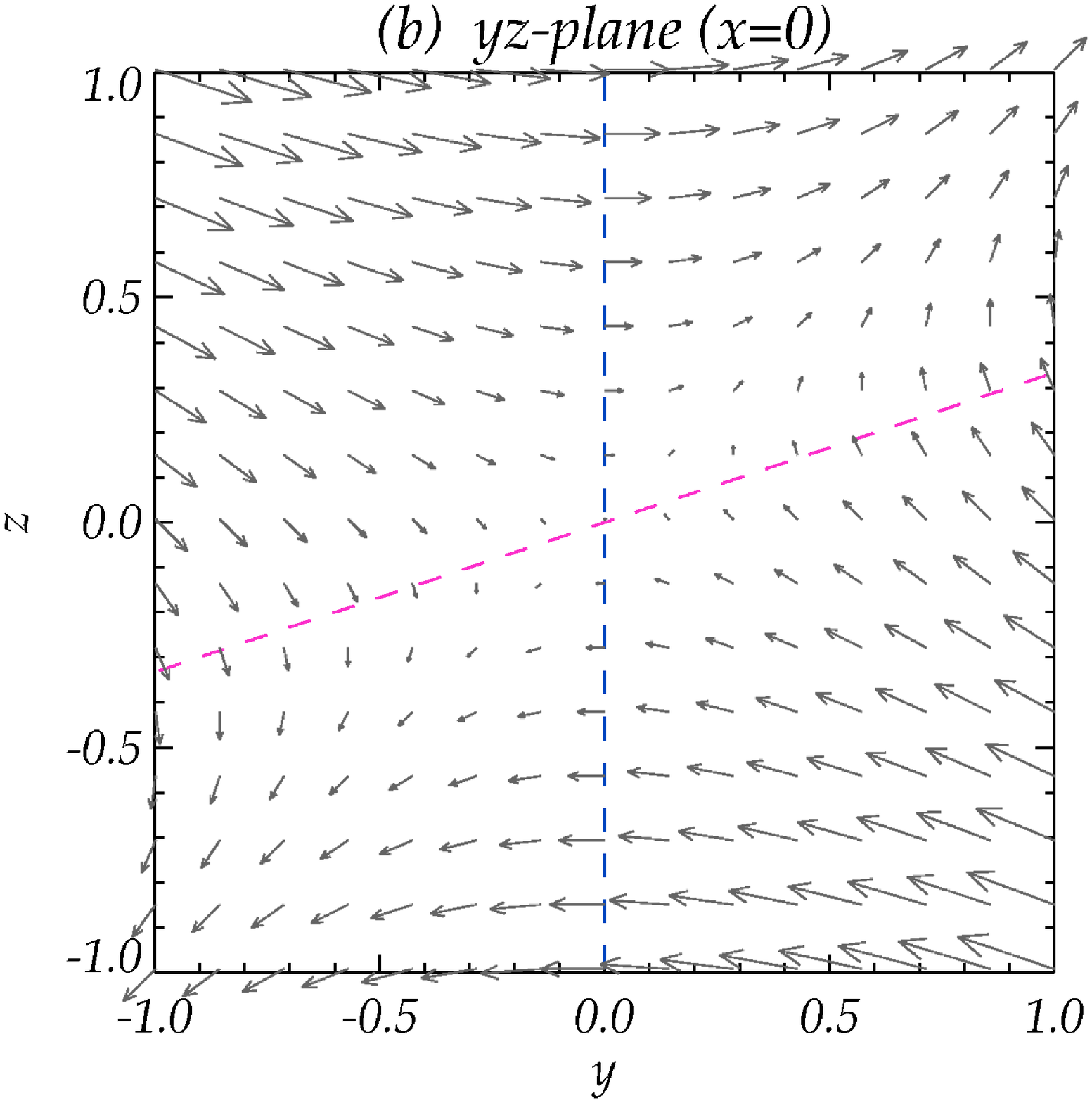}
    \end{minipage}

    \caption{Magnetic configuration for the initial non-equilibrium state with homogeneous fan-aligned current, for the case with $j_{f}=1$ and $p_0=1$, showing (a) the 3D configuration with field lines above and below the fan in purple and orange, respectively. The fan plane is outlined in dashed black, and the spine is represented in green, with projections onto the $xz$-plane and $yz$-plane (dashed green lines). In (b), the spine (blue) and the section of the fan plane (pink) are plotted in the $x=0$ plane, and the grey arrows show the direction of the initial Lorentz forces.}
    \label{fig:fan_initial}
  \end{minipage}
  \vspace{0.3cm}

  \begin{minipage}[b]{1.0\linewidth}

    \begin{minipage}[b]{0.50\linewidth}
      \centering
      \includegraphics[scale=0.32]{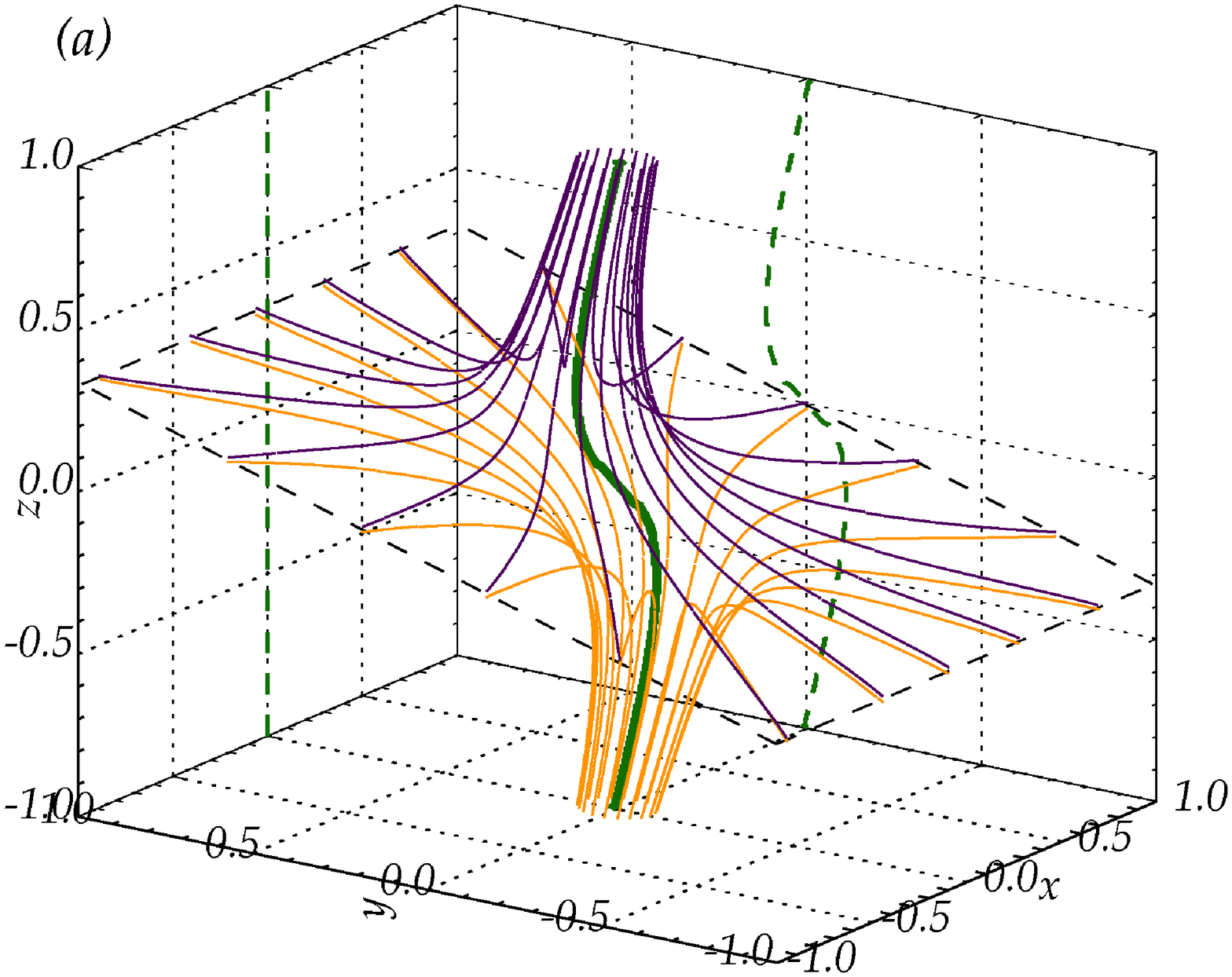}
    \end{minipage}
    \begin{minipage}[b]{0.50\linewidth}
      \centering
      \includegraphics[scale=0.30]{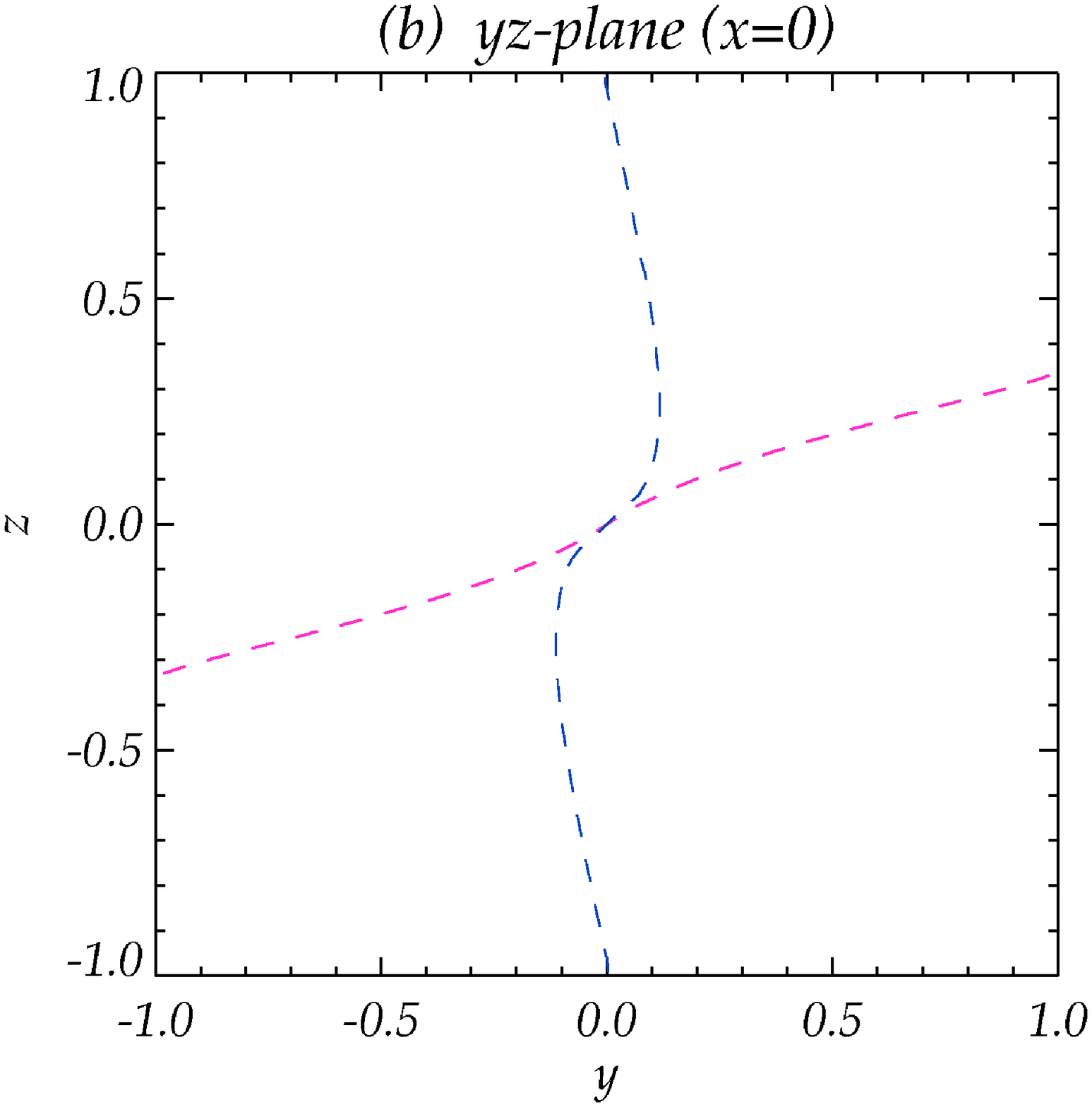}
    \end{minipage}

    \caption{Magnetic configuration for the final equilibrium state for the radial tilted null shown in Fig. \ref{fig:fan_initial}, showing (a) the 3D configuration with field lines above and below the fan in purple and orange, respectively. The fan plane is outlined in dashed black and the spine is represented in green, with projections onto the $xz$-plane and $yz$-plane in dashed green lines. In (b), the spine (blue) and the section of the fan plane (pink) are plotted in the $x=0$ plane.}
    \label{fig:fan_final}
  \end{minipage}
\end{figure*}

As in \citet{Fuentes12c}, for the numerical experiments we used Lare3D, a staggered Lagrangian-remap code that solves the full MHD equations with user-controlled viscosity \citep[see][]{Arber01}. The resistivity is set to zero for our experiments, and the numerical domain is a 3D uniform grid of $512^3$ points. The boundary conditions are closed, the magnetic field lines are line-tied, and the MHD equations are normalised following a standard procedure. For further details on the boundary conditions and normalisation, see Sec. 3 of \citet{Fuentes12c}.

Thus, the normalised ideal MHD equations are as follows:
\begin{eqnarray}
\frac{\partial \rho}{\partial t}+\boldnabla\cdot(\rho{\bf v}) &=& 0\;,\label{n_mass}\\
\rho\frac{\partial{\bf v}}{\partial t}+\rho({\bf v}\cdot\boldnabla){\bf v} &=& -\boldnabla p + (\boldnabla\times{\bf B})\times{\bf B} + {\bf F}_{\nu}\;,\label{n_motion}\\
\frac{\partial p}{\partial t}+{\bf v}\cdot\boldnabla p &=& -\gamma p \boldnabla\cdot{\bf v}+H_{\nu}\;,\label{n_energy}\\
\frac{\partial{\bf B}}{\partial t} &=& \boldnabla\times({\bf v}\times{\bf B})\;,\label{n_induction}
\end{eqnarray}
where ${\bf F}_{\nu}$ and $H_{\nu}$ are the viscous force and viscous heating terms, and the internal energy, $\epsilon$, is given by the ideal gas law, $p=\rho\epsilon(\gamma-1)$, with $\gamma=5/3$. 


\section{Fan-parallel current density} \label{sec4}


\subsection{Initial state}

We first look at the relaxation of initial configurations of a magnetic null point with constant current density everywhere in the direction parallel to the fan surface, but perpendicular to the spine (i.e., along the $x$-axis), of the form $(j_{f},0,0)$. For simplicity, we assume $b=1$ in Eq. (\ref{M_general}), so the field lines have no preferred direction on the fan plane, but expand regularly through the plane.

\begin{figure*}[t]
  \begin{minipage}[b]{1.0\linewidth}
    \centering
    \includegraphics[scale=0.60]{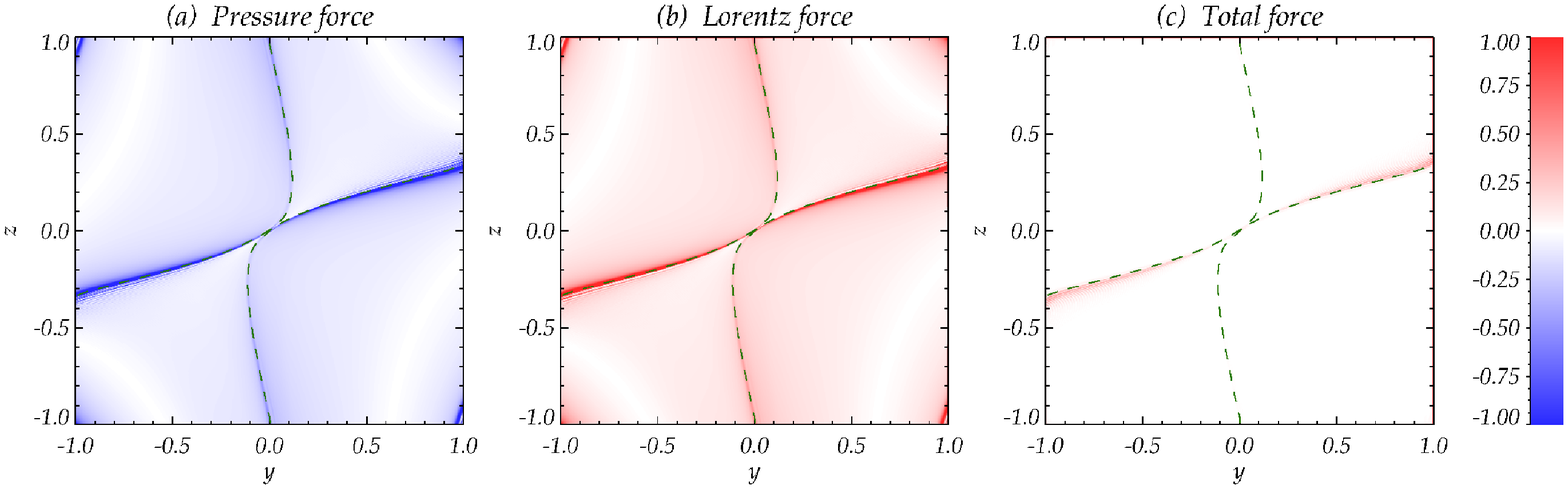}
    \caption{Contour plots of the different forces acting in the final equilibrium state in the $x=0$ plane, for the same experiment as in Fig. \ref{fig:fan_initial}. Showing, from left to right, the magnitude of the pressure force ($-|\boldnabla p|$), of the Lorentz force and of the total force. Values are normalised to the maximum force of the initial state. It can be observed that the pressure and Lorentz forces balance each other out, creating a clearly non-force-free equilibrium.}
    \label{fig:fan_forces}
  \end{minipage}
  \vspace{0.3cm}

  \begin{minipage}[b]{1.0\linewidth}

    \begin{minipage}[b]{0.49\linewidth}
      \centering
      \includegraphics[scale=0.35]{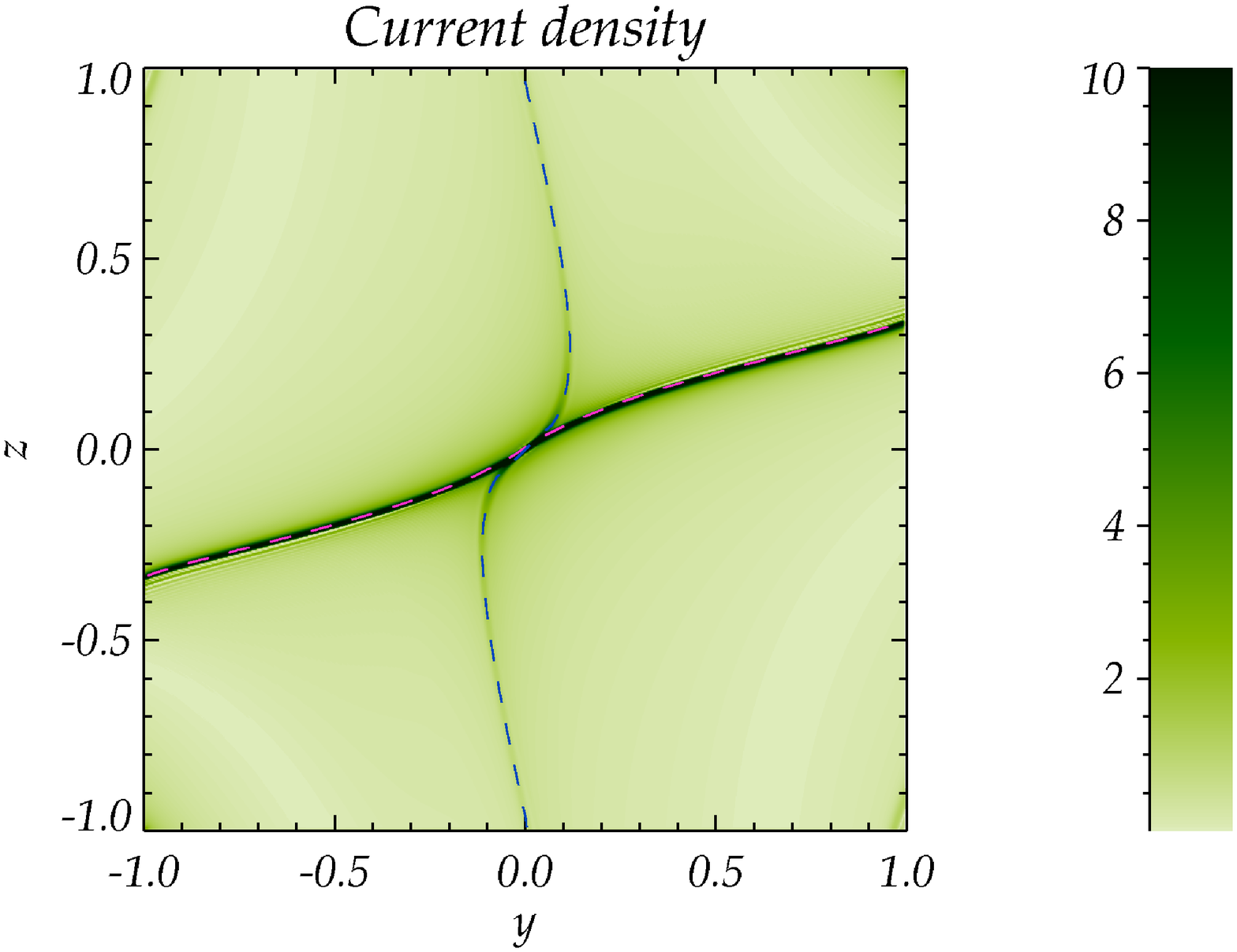}
    \end{minipage}
    \hspace{0.02\linewidth}
    \begin{minipage}[b]{0.49\linewidth}
      \centering
      \includegraphics[scale=0.32]{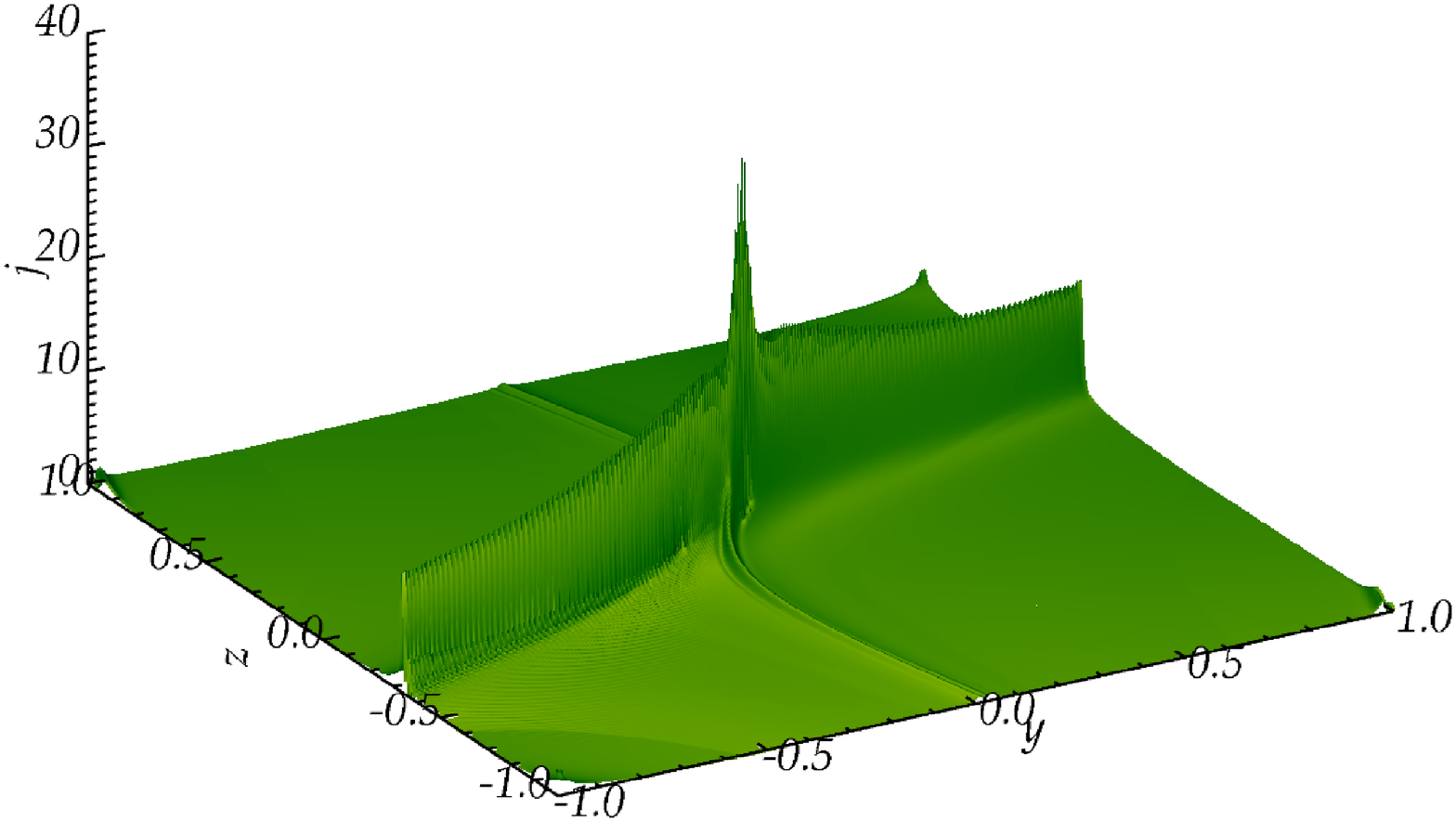}
    \end{minipage}

    \caption{Contour plot (left) and corresponding surface (right) of the magnitude of the electric current density in the final equilibrium state, for the same experiment as in Fig. \ref{fig:fan_initial}, in a cross section in the $x=0$ plane. }
    \label{fig:fan_current}  
  \end{minipage}
\end{figure*}

Our initial magnetic field is then given by
\begin{equation}
(B_x, B_y, B_z)=(x, y, j_f\,y-2z)\;.
\end{equation}
Initially, the spine line lies along the $z$-axis, and the fan plane is not perpendicular to the spine (unlike in \citet{Fuentes12c}). Instead, it is tilted about the $x$-axis (the direction of the current density), and lies in the plane
\begin{equation}
z=\frac{j_f}{3}y\;,
\end{equation}
\citep{Parnell96}. We ran various experiments with different initial plasma pressures and current densities. Figure \ref{fig:fan_initial} shows the magnetic configuration of the initial state, for $j_{f}=1$ and $p_0=1$.

Initially, for such a non-potential null, the Lorentz force, ${\bf j}\times{\bf B}$, has the $y$ and $z$ components that are non-zero. The forces in the initial state act are shown in Fig. \ref{fig:fan_initial}b, pushing the spine to the right above the fan and to the left below it, and pushing the fan up for positive values of $y$ and down for negative values of $y$. Therefore, the spine and fan will collapse into each other in the direction of the original tilt of the fan.


\subsection{Final equilibrium state}

We initially concentrate on the case shown in Fig. \ref{fig:fan_initial}, with $j_f=1$ and $p_0=1$. The numerical simulations finish at about 300 fast magnetosonic travel times (defined as the time for a fast magnetosonic wave to travel from the null to one of the boundaries). The magnetic field configuration in the final state shown in Fig. \ref{fig:fan_final}. In comparison to the initial state (Fig. \ref{fig:fan_initial}), the relaxation collapses the spine towards the fan in the region near the null, but the fan plane is only slightly perturbed. The total energy of the evolution is checked to be conserved throughout the relaxation to within an error of 0.02\%. This conservation of energy demonstrates that numerical diffusion does not play a significant role in the relaxation. Less than 2\% of the initial magnetic energy is converted to internal energy via viscous damping during the relaxation. 

The final state is a non-force-free equilibrium where the non-zero Lorentz forces are globally balanced by non-zero plasma pressure forces. In Fig. \ref{fig:fan_forces}, we show the different forces in the vertical $x=0$ plane. In this experiment, the system has reached a non-force-free equilibrium with the strongest pressure force and Lorentz force concentrated around the spine line and about the fan plane, and the field is close to force-free everywhere else. Weak residual total forces remain near the fan away from the null. These forces are the consequence of the line tiding of the field (of the fan plane in particular) at the boundaries, and they would take much longer to completely disappear. Nonetheless, they do not influence the results of the present study.

\begin{figure}[t]
  \centering
  \includegraphics[scale=0.32]{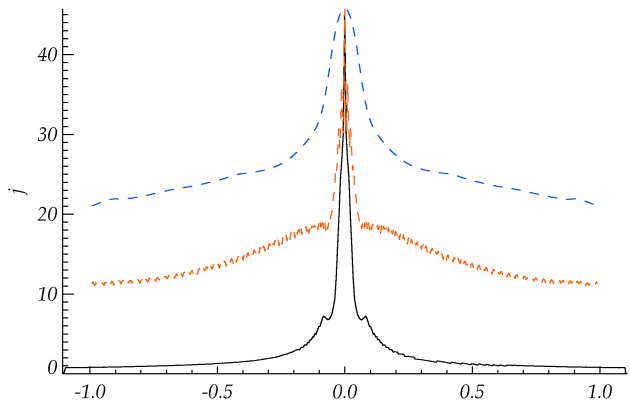}
  \caption{Plots of the magnitude of the current density for three different cuts, along the spine line (black), along the tilt-axis ($x$-axis) of the fan (dashed blue) and along the $y$-axis on the fan surface (dashed orange).}
  \label{fig:fan_along}
\end{figure}
\begin{figure}[t]
  \centering
  \includegraphics[scale=0.35]{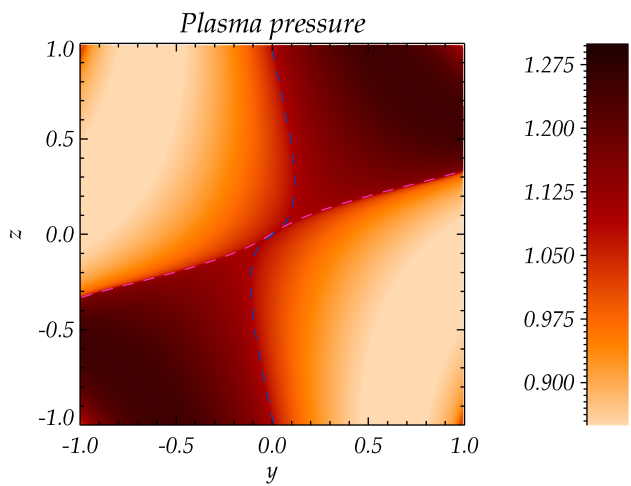}
  \caption{Contour plot of the equilibrium plasma pressure for the same experiment as in Fig. \ref{fig:fan_initial}, for a vertical cross section in the $x=0$ plane.}
  \label{fig:fan_pressure}
\end{figure}

The final distribution of the magnitude of the current density in the $x=0$ plane is shown in Fig. \ref{fig:fan_current}. Not surprisingly, the current accumulations appear at the same locations as the strongest pressure forces and Lorentz forces shown in Fig. \ref{fig:fan_forces}, namely, on the fan plane and along the spine. The current density is approximately zero away from the fan and the spine. It shows a weak accumulation along the spine line, but is about one order of magnitude stronger on the fan plane. At the location of the null we find a pronounced peak of the current density. The whole picture is reminiscent of the 2D X-point collapse \citep{Fuentes11}, in which the current accumulates along the four separatrices and the system evolves towards an infinite time singularity at the location of the null (but contrasts with \citet{Fuentes12c} where the current accumulates in the vicinity of the spines). \citet{Parnell97} proved that the magnetic field locally about a non-potential 3D null point, i.e. the linear field about a non-potential null, produces a Lorentz force that cannot be balanced by a plasma pressure force, and so there are no static equilibrium models of linear non-potential or force-free nulls. Therefore, even if the system is in global equilibrium in the final state, the collapse of the fan and spine towards each other does not allow a potential configuration at the null, and locally, the pressure forces cannot balance the Lorentz forces there, which gives rise to the formation of a singularity. Unfortunately, the same analysis of forces about the null defined in \citet{Fuentes11} to evaluate the formation of the singularity cannot be made here, due to the relatively low grid resolution that places the small local region in which these forces act into just a few gridpoints. The formation of a continuously growing singularity at the location of the null is evaluated in Sec. \ref{sec:sing}.

The final current layer around the location of the null can be better appreciated in Fig. \ref{fig:fan_along}. Here, we show three different cuts of the current density, along the line of the spine, along the tilt axis of the fan surface ($x$-axis), and along the $y$-axis of the fan surface. We see that the current density on the fan plane is enhanced over ten times that of the initial background current density, and it is higher in a thin layer along the $x$-axis, which is the initial tilt axis of the fan. The amplitude of the current layer is greatest at the location of the null, corresponding to the pronounced peak observed in Fig. \ref{fig:fan_current}. The distribution of current density observed in Fig. \ref{fig:fan_along} agree qualitatively with the results of \citet{Pontin05b}, who consider the collapse of actually the same initial field, using a different numerical method. The comparison shows that the way the current density accumulates about the null is not an artefact of the numerical method, but a physical result.

In Fig. \ref{fig:fan_pressure}, we show the final distribution of plasma pressure in the $x=0$ plane. The plasma pressure is enhanced inside the regions where the spine and fan have collapsed towards one another, but decreased in the regions outside. The highest residual gradients in pressure occur at the locations of the fan and spine, and the gradient of plasma pressure from positive to negative y, far from the null point (or far from the spine line in general), is not as sharp as it is at the x=0 plane (the one in the figure). Once again, this result brings out the fact that the collapse of a 3D null points resemble the 2D collapse in different ways.


\subsection{Singular current} \label{sec:sing}

Finally, for this setup, we want to evaluate the formation of the singularity at the location of the null, by looking at the time evolution of the maximum current, at the location of the null, for experiments with different initial plasma pressures and different initial currents (inclinations of the fan plane). In particular, we want to see if there is a current singularity being formed, and if there is, whether its growth rate implies an infinite or finite rate of formation.

\citet{Klapper97} rigorously demonstrated that, for 2D ideal incompressible plasmas, a singularity of the current density will take an infinite amount of time to develop, unless driven by a pressure singularity occurring outside the neighbourhood of the null point. \citet{Fuentes11} showed numerically that this is also the case for the 2D compressible collapse of magnetic X-points, and here, we aim to extend these results for 3D nulls with fan-aligned current.

\begin{figure}[t]
  \includegraphics[scale=0.32]{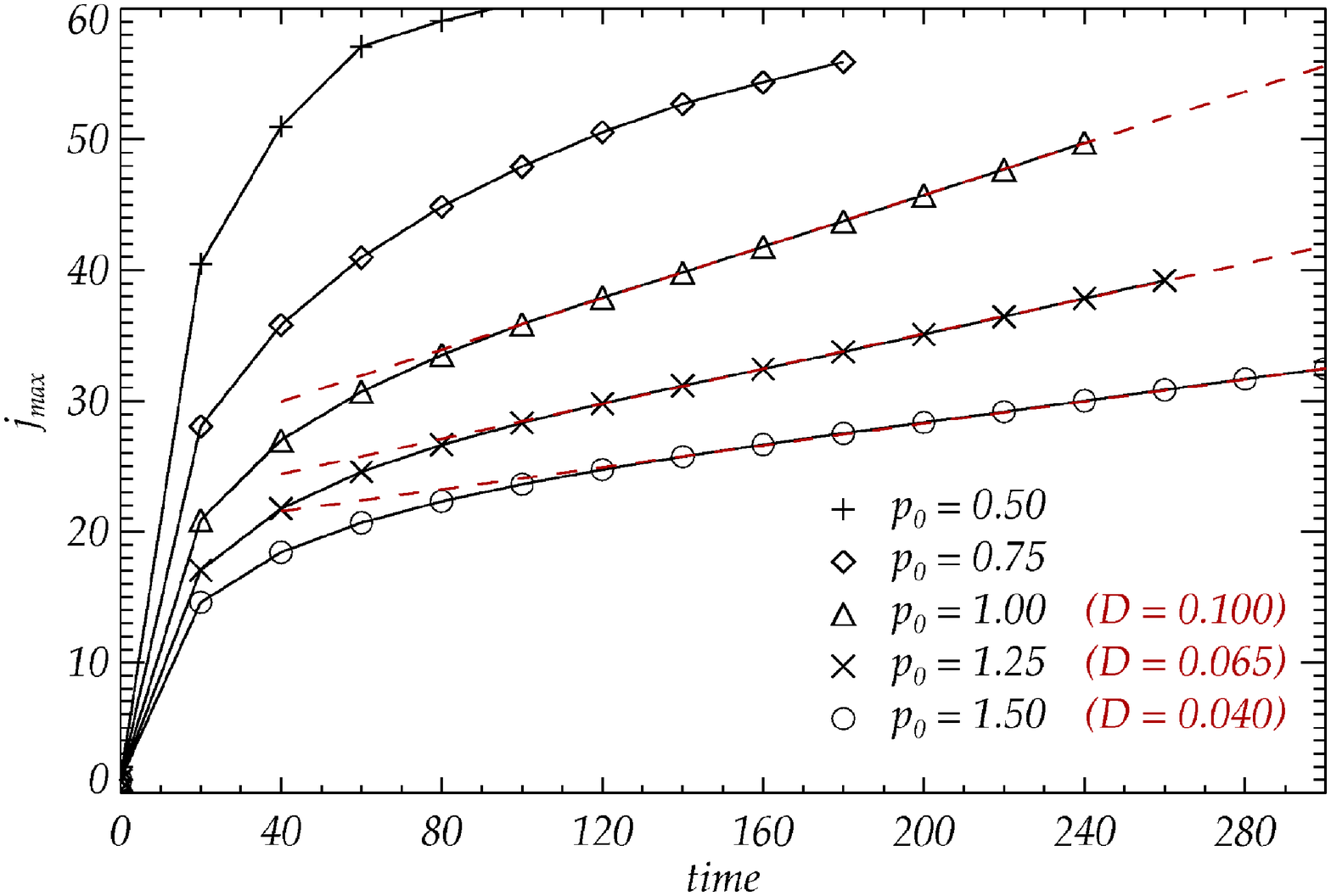}
  \caption{Time evolution of the peak current for five experiments with different initial plasma pressures and an initial current of $j_f=1$. When a good asymptotic regime is achieved, a fit of $j_{max}=C+Dt$ is overplotted (red dashed line).}
  \label{fig:fan_peak}
  \vspace{0.3cm}

  \includegraphics[scale=0.32]{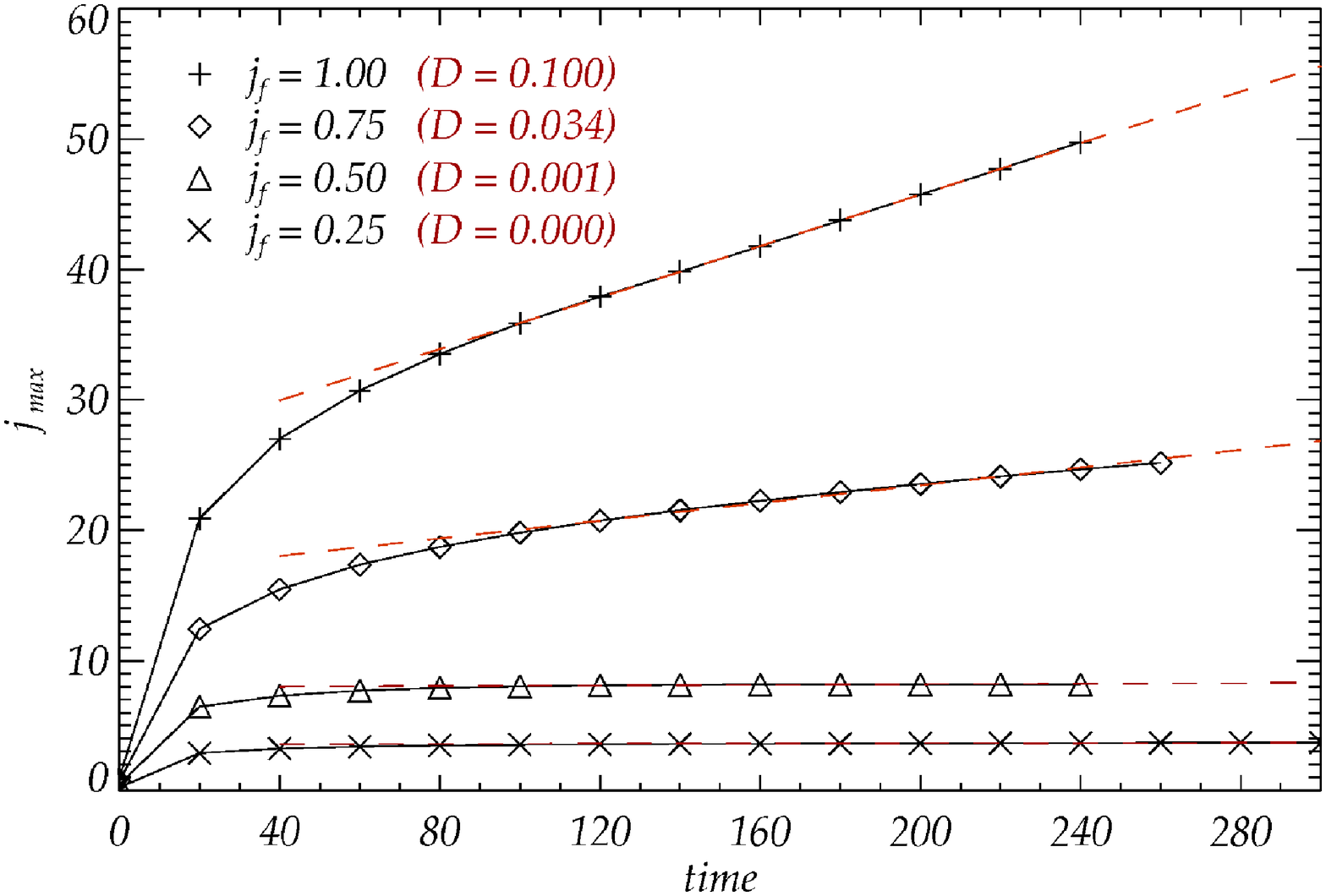}
  \caption{The same as Fig.~\ref{fig:fan_peak}, but for four experiments with different initial current densities and an initial plasma pressure of $p_0=1$. }
  \label{fig:fan_peak2}
\end{figure}

The ideal MHD relaxation of a 3D null point with an initial homogeneous current density parallel to the fan plane causing (or caused by) a tilt of the fan, results in the collapse of the fan and the spine towards each other. The current density accumulates weakly along the spine and on the fan plane forming a ridge of current along the tilt axis of the fan (axis of the initial current) which spikes at the null itself. The continuous concentration of current at the location of the null is a natural consequence from the results of \citep{Parnell97}, according to which the Lorentz forces cannot be balanced by plasma pressure forces about the 3D null. These forces feed current at the locations of the null point slowly, forming an infinite time singularity. Figures \ref{fig:fan_peak} and \ref{fig:fan_peak2} show the time evolution of this peak current for varying initial plasma pressures and current densities, respectively. In Fig. \ref{fig:fan_peak}, the current is set to $j_f=1$ and the pressure varies from $p_0=0.5$ to $p_0=1.5$. After the rapid viscous relaxation, the system enters a regime in which the maximum current follows a linear function of the form $j_{max}=C+Dt$, where the growth rate of the singularity, $D$, decreases as the initial plasma pressure, $p_0$, increases, as was found in the compressible 2D null collapse experiments of \citet{Fuentes11}. For the experiments with the lowest initial plasma pressures, a fit of this form cannot be made, as the numerical diffusion limit is reached before a well-defined linear regime has been achieved.

Similar linear behaviour is found in Fig. \ref{fig:fan_peak2}, where the pressure is set to the constant value $p_0=1$, and the current density is varied from $j_f=0.25$ to $j_f=1$. When the initial current of the system is decreased then naturally the current in the singularity also decreases as does the rate at which the singularity forms (Fig.~\ref{fig:fan_peak2}). A higher initial current density produces a bigger collapse of the fan and spine towards each other, hence an increase in the magnitude and growth rate of the singularity. Indeed, when the initial current is particularly weak then the collapse is practically negligible, so it is the rate of growth of the singularity.

\begin{figure}[t]
  \centering
  \includegraphics[scale=0.32]{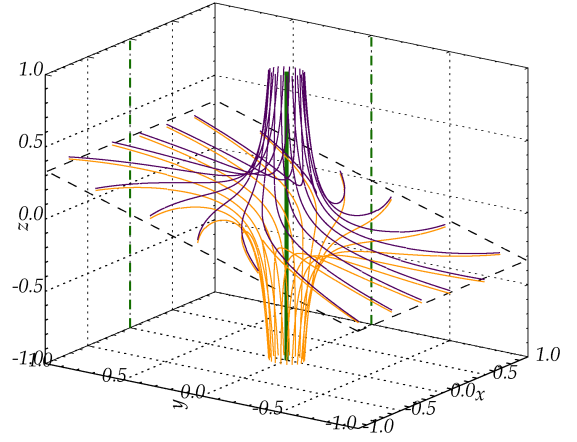}
  \caption{3D magnetic configuration of the initial non-equilibrium state with generic current density, for the case with $j_{sp}=j_f=1$ and $p_0=1$, showing the same features as in Fig. \ref{fig:fan_initial}.}
  \label{fig:generic_initial}
  \vspace{0.3cm}
\end{figure}


\section{Generic current density} \label{sec5}

In the previous section, we considered the MHD relaxation of an initial radial null with a component current in the fan plane, perpendicular to the spine. In such an initial field the fan plane is tilted with the current aligned with the axis of tilt. In this section, we consider the numerical equilibrium resulting from the MHD relaxation of an initial radial null that has a generic current, with non-zero components along both the spine and the fan. To our knowledge the collapse of these types of nulls have not been considered before. Here, the magnetic field of the initial null consists of a tilted fan, as in the previous case, but this time with twisted field lines around the spine. \citet{Fuentes12c} considered the dynamical relaxation of a null with only a component of current parallel to the spine, which gives rise to twisted field about the spine, but maintains the fan and spine at right angles.


\subsection{Initial state}

A 3D null configuration of a radial null with a generic current density of the form ${\bf j}=\frac{1}{\mu}(j_f, 0, j_{sp})$ is considered. From Eq. (\ref{M_general}), the magnetic field of the initial state is given by
\begin{equation}
(B_x, B_y, B_z)=(x-\frac{j_{sp}}{2}y,\; \frac{j_{sp}}{2}x+y,\; j_f\,y-2z)\;.
\end{equation}
Here again we have assumed $b=1$ for simplicity. For the case studied here we have chosen the background plasma pressure to be $p_0=1$ and, for the current density, $j_{sp}=j_f=1$. Thus, in the initial state, the current density vector lies in the $xz$-plane, at $45^{\circ}$ to both the $x$ and the $z$ axis, and its magnitude is $|j|=\sqrt{2}$ everywhere. Initially, the spine line lies along the $z$-axis, the fan plane tilts about the $x$-axis following the function $z=\frac{j_f}{3}y$, and the magnetic field lines show a homogeneous twist about the spine (see Fig. \ref{fig:generic_initial}). The Lorentz forces that act in the system initially are a combination of two contributions. The first is the one shown in Fig. \ref{fig:fan_initial}b, which will drive the collapse of the fan and spine towards one another. The second is a magnetic tension force, as described in \citet{Fuentes12c}, which will remove the twist from the fan surface.


\subsection{Final equilibrium state}

\begin{figure}[t]
  \centering
  \includegraphics[scale=0.32]{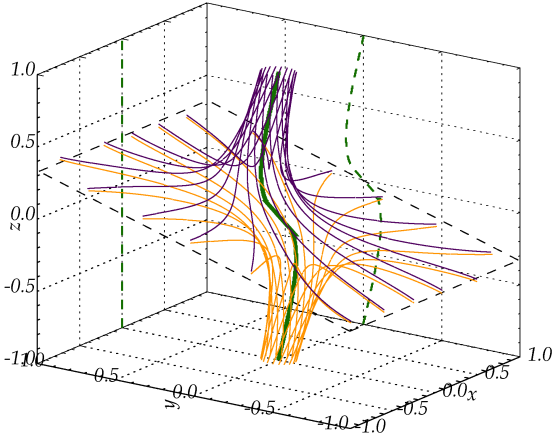}
  \caption{3D magnetic configuration for the final equilibrium with generic current density, for the same experiment as in Fig. \ref{fig:generic_initial}, showing the same features as in Fig. \ref{fig:fan_initial}.}
  \label{fig:generic_final}
 \end{figure}

\begin{figure}[t]
  \centering
  \includegraphics[scale=0.35]{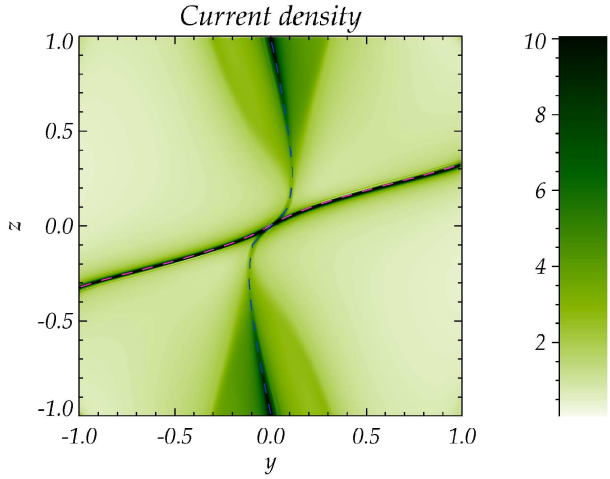}
  \caption{Contour plot of the magnitude of the electric current density in the final equilibrium state, for the same experiment as in Fig. \ref{fig:generic_initial}, in a cross section in the $x=0$ plane.}
  \label{fig:gen_current}
\end{figure}

The results of the ideal relaxation from this initial configuration shown in Fig. \ref{fig:generic_initial} combine the features from the relaxation of spiral nulls \citep{Fuentes12c} and those of tilted nulls, from Sec \ref{sec4}. The magnetic field lines in the final equilibrium state are shown in Fig. \ref{fig:generic_final}. The relaxation drives the collapse of the fan and the spine towards each other at the same time as it concentrates the initial twist of the field lines about the spine line, above and below the fan surface. This can be more clearly seen in Fig. \ref{fig:gen_current}, where we plot the magnitude of the current density for the final equilibrium state, in the $x=0$ plane. It concentrates in three different regions: (i) on the fan surface, due to the spine-fan collapse; (ii) around the spine above the fan; and (iii) around the spine below the fan. Clearly, accumulation (i) is the same as what is seen in Fig. \ref{fig:fan_current} and is therefore associated with the component of current perpendicular to the spine, whereas the last two accumulations result from a concentration of the initial twist of the field lines about the spine and resemble the hourglass configuration obtained in \citet{Fuentes12c}. They are thus associated with the component of current parallel to the spine.

\begin{figure}[t]
  \includegraphics[scale=0.32]{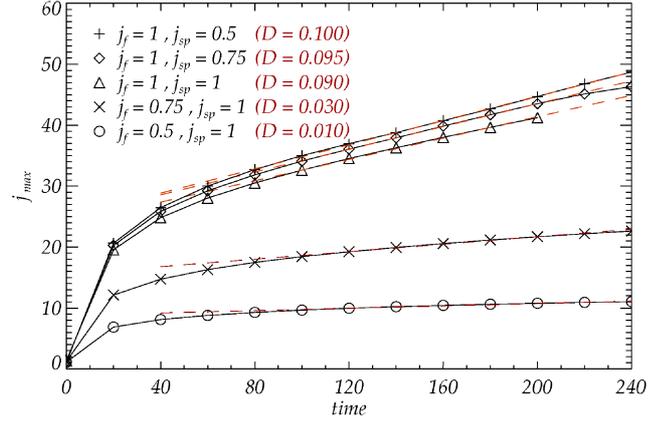}
  \caption{Time evolution of the peak current for five experiments with different initial current densities involving $j_f$ and $j_{sp}$, and an initial plasma pressure of $p_0=1$. When a good asymptotic regime is achieved, a fit of $j_{max}=C+Dt$ is overplotted (red dashed).}
  \label{fig:fanspine_peak}
\end{figure}

We now evaluate the consequences of the two deformations (twist and tilt) on the formation and growth rate of the singularity at the null. In Fig. \ref{fig:fanspine_peak} we show the time evolution of the peak current density (the current density at the null) for five different experiments with a fixed background plasma pressure ($p_0=1$) and varying both $j_f$ and $j_{sp}$ independently. When $j_{sp}$ is varied and $j_f$ is fixed, the growth rate of the singularity does not vary significantly, as the torsional disturbance leads to a current accumulation about the spines, not at the null. However, a decrease in $j_f$ for fixed {\bf $j_{sp}$} slows down the formation of the singularity, as discussed below.

When the initial current density has the form ${\bf j}=(j_f,0,j_{sp})$, where $j_{sp}=j_f=1$, both contributions of current density initially parallel and perpendicular to the spine are the same as in each of these independent cases studied before: the case $j_f=1$ and $j_{sp}=0$ studied previously in this paper, and the case $j_f=0$ and $j_{sp}=1$ studied in \citet{Fuentes12c}. The magnitude and growth rate of the singularity for $j_f=1$ and $j_{sp}=1$ ($D=0.09$) is to be compared with the results from Sec. \ref{sec:sing} for the case $j_f=1$ and $j_{sp}=0$ ($D=0.1$). The addition of a spine-aligned component of the current density slows down the formation of the singularity.

\begin{figure}
  \centering
  \includegraphics[scale=0.1]{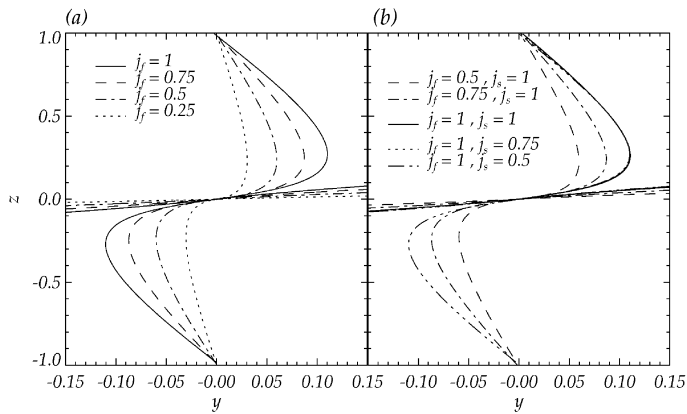}
  \caption{Plots of the $yz$-plane at $x=0$ showing the spine and the cut along the fan plane, for (a) the experiments with fan parallel current density only and (b) the experiments with generic current density. In both plots, the non-zero components of the initial current density are varied systematically.}
  \label{fig:yzcuts}
\end{figure}

The decrease in the growth rate of the singularity may be explained as follows. The fan-aligned component of the current density, $j_f$, curves the spine line making it longer within the boundaries of the domain. This way, the relaxation of the twist component of the magnetic field (identified with the spine-aligned component of the current density) requires more current to be allocated along and around the spine than in the absence of $j_f$ (when the spine line is straight), thereby decreasing the amount of current that is available for the formation of the singularity.

The curvature of the spine line may be appreciated in Fig. \ref{fig:yzcuts}. Here we show plots of the $yz$-plane at $x=0$, indicating the spines and the cuts along the fan plane for different experiments. Figure \ref{fig:yzcuts}a corresponds to the experiments of Sec.~\ref{sec4} with fan-aligned current density. Clearly, as $j_f$ increases, the spines become more curved and get longer. Figure \ref{fig:yzcuts}b corresponds to the experiments discussed in the present section with generic current density. We note that changes in $j_{sp}$, for fixed $j_f$, do not involve a change in the shape of the spine line or in the fan plane, i.e. they all lie under the solid lines of the $j_{sp}=j_f=1$ case. On the contrary, increasing values of $j_f$, for fixed $j_{sp}$, lead to larger deformations of the spine and the fan, hence of the current layer about the null. This is due to the greater initial inclination of the fan plane that, upon the ideal MHD relaxation of the system, is transferred into a greater folding of the spine towards the fan plane. 

Over all, when a current perpendicular and parallel to the spine are combined together, the resulting non-force-free equilibrium is such that it combines the results from a strictly tilted null with that from a strictly spiral null. Te spiraling of the field lines along the spine does not concentrate any current at the location of the null \citep{Fuentes12c}, so it is not a direct contribution to the formation of the singularity. In contrast, the addition of a current density parallel to the spine is found to slow down the formation of the singularity itself.


\section{Summary and conclusions} \label{sec7}

The 3D relaxation of tilted magnetic null points with an initial non-zero component of the current density perpendicular to the spine has been investigated under non-resistive conditions, in all cases resulting in a non-force-free equilibrium, where all the velocities have been damped out by viscous forces. In the final equilibrium states, the forces due to the plasma pressure and the magnetic field, which are of about the same strength as the initial non-equilibrium Lorentz forces, balance each other out, resulting in a genuine non-force-free equilibrium, save at the location of the null where an infinite time singularity occurs, similar to the 2D X-point collapse studied by \citet{Fuentes11}.

During the relaxation, for the case with an initial radial null and constant current density purely parallel to the fan (i.e., a tilted fan on which field lines expand radially), the field evolves by warping the fan and the spine so that, locally about the null, they collapse towards each other, concentrating the current density on the fan surface, but leaving it weak along the spine. The amplitude of the resulting current density on the fan is largest along the $x$-axis, the initial tilt axis of the fan along which the initial and final current densities are directed, and it shows a sharp peak at the location of the null. 

After the viscous relaxation has finished, the system has entered an asymptotic regime in which the current at the location of the null increases linearly with time. This behaviour is identified with the slow formation of an infinite time singularity whose growth rate depends directly on the values of the initial plasma pressure and current density (tilt of the fan). A decrease in the plasma pressure at a fixed current density or an increase in the electric current density at a fixed plasma pressure produce an increase in the growth rate of the singularity being formed at the location of the null.

Overall, the structure of the equilibrium is very similar to the 2D X-point collapse studied in \citet{Fuentes11}; in particular, (i) the current density accumulations along the fan and spine are equivalent to the accumulations found along the separatrices in the 2D X-point case, (ii) the collapse and deformation of these features towards each other locally about the null is observed in both cases, as is (iii) the formation of an infinite time singularity at the null. The results also agree with past studies of compressible 3D nulls with current purely perpendicular to the spine that collapse to form current singularities \citep{Pontin05b}. However, we have been able to study the exact distribution of the current in the final equilibrium state in more detail. Furthermore, our numerical code allows us to determine the full dynamical evolution of the system, and therefore we are able to get a numerical growth rate for the singularity, which directly depends on the initial plasma pressure. 

In \citet{Fuentes12c} where the initial magnetic field was twisted about the spine owing to current being parallel to the spine (along the $z$-axis), (i) the spine and fan remained perpendicular and did not collapse, (ii) the current (which remained parallel to the spine) accumulated in a pair of cone shape features about the spine, and (iii) no current singularity was found at the null in the radial null case. Thus the current accumulations are totally different in these two scenarios, as expected. 

The effects of a more generic current with components along both the fan and spine have not been considered before the present paper. In the cases where the initial current density has both components parallel and perpendicular to the spine, the initial field shows a tilted fan on which field lines expand in spirals around the null. The resulting non-force-free equilibrium combines the two contributions, from the tilted fan and the twisted field lines, creating a current density that concentrates on the fan plane, building a singularity at the location of the null, and along the spine line, concentrating the twist of the field lines there. In principle, the results are a simple combination of the twisted and tilted cases, and so, the current density should be evenly ditributed for both contributions, even though the magnitude and growth rate of the singularity has been decreased. This is because, since the spine is longer now than in the $j_f=0$ and $j_{sp}=1$ case (because it is bended), the relaxation of the twist requires a bit more current, thereby slightly decreasing the amount of current available for the singularity at the null. The addition of a spine-aligned component of the current density slows down the formation of the singularity.

In every case considered here, we set $b=1$ in the magnetic field. This means that the two eigenvalues that are associated with the fan plane have the same magnitude, and thus, the fan plane has no major and minor axes, so the field lines have no preferred direction in the fan plane \citep{Parnell96}. The presence of major or minor axes in a fan plane may well have an effect on the equilibria obtained following the collapse of such a null with a non-zero component of current.   

Magnetic reconnection is an important process of energy release in scenarios like the solar corona and the Earth's magnetotail. In the second case, it provides a mechanism for particle acceleration and for the global aurora. In the first case, it is the mechanism for particle acceleration, solar flares, and coronal mass ejections, and is highly likely to provide an important source of energy for the high temperatures observed in the corona. In the present paper, we have described a valid initial equilibrium field for the study of spontaneous magnetic reconnection, initiated by microinstabilities at a 3D magnetic null. Such a study is a natural continuation of the work carried out by \citet{Fuentes12a} and \citet{Fuentes12b} on spontaneous reconnection and the onset of impulsive bursty reconnection at non-force-free current layers.


\section*{Acknowledgements}

The authors would like to thank the referee for many useful comments that helped us to clarify the details of the experiments carried out in this paper. JFF gratefully acknowledges funding from the St Andrews Rolling Grant (ST/H001964/1). This work was partly supported by SOLAIRE European training network. Computations were carried out on the UKMHD consortium cluster funded by the STFC and SRIF.


\bibliographystyle{aa}
\bibliography{20346}

\begin{thebibliography}{48}
\expandafter\ifx\csname natexlab\endcsname\relax\def\natexlab#1{#1}\fi

\bibitem[{{Al-Hachami} \& {Pontin}(2010)}]{Alhachami10}
{Al-Hachami}, A.~K. \& {Pontin}, D.~I. 2010, \aap, 512, A84

\bibitem[{{Arber} {et~al.}(2001){Arber}, {Longbottom}, {Gerrard}, \&
  {Milne}}]{Arber01}
{Arber}, T.~D., {Longbottom}, A.~W., {Gerrard}, C.~L., \& {Milne}, A.~M. 2001,
  Journal of Computational Physics, 171, 151

\bibitem[{{Aulanier} {et~al.}(2006){Aulanier}, {Pariat}, {D{\'e}moulin}, \&
  {Devore}}]{Aulanier06}
{Aulanier}, G., {Pariat}, E., {D{\'e}moulin}, P., \& {Devore}, C.~R. 2006,
  \solphys, 238, 347

\bibitem[{{Bulanov} {et~al.}(2002){Bulanov}, {Echkina}, {Inovenkov}, \&
  {Pegoraro}}]{Bulanov02}
{Bulanov}, S.~V., {Echkina}, E.~Y., {Inovenkov}, I.~N., \& {Pegoraro}, F. 2002,
  Physics of Plasmas, 9, 3835

\bibitem[{{Cowley}(1973)}]{Cowley73}
{Cowley}, S.~W.~H. 1973, Cosmic Electrodyn., Vol.~3, p.~448 - 501, 3, 448

\bibitem[{{Craig} \& {Litvinenko}(2005)}]{Craig05}
{Craig}, I.~J.~D. \& {Litvinenko}, Y.~E. 2005, Physics of Plasmas, 12

\bibitem[{{Demoulin} {et~al.}(1997){Demoulin}, {Bagala}, {Mandrini}, {Henoux},
  \& {Rovira}}]{Demoulin97}
{Demoulin}, P., {Bagala}, L.~G., {Mandrini}, C.~H., {Henoux}, J.~C., \&
  {Rovira}, M.~G. 1997, \aap, 325, 305

\bibitem[{{Demoulin} {et~al.}(1996){Demoulin}, {Henoux}, {Priest}, \&
  {Mandrini}}]{Demoulin96}
{Demoulin}, P., {Henoux}, J.~C., {Priest}, E.~R., \& {Mandrini}, C.~H. 1996,
  \aap, 308, 643

\bibitem[{{Edmondson} {et~al.}(2009){Edmondson}, {Lynch}, {Antiochos}, {De
  Vore}, \& {Zurbuchen}}]{Edmondson09}
{Edmondson}, J.~K., {Lynch}, B.~J., {Antiochos}, S.~K., {De Vore}, C.~R., \&
  {Zurbuchen}, T.~H. 2009, \apj, 707, 1427

\bibitem[{{Fuentes-Fern{\'a}ndez} \& {Parnell}(2012)}]{Fuentes12c}
{Fuentes-Fern{\'a}ndez}, J. \& {Parnell}, C.~E. 2012, \aap, 544, A77

\bibitem[{{Fuentes-Fern{\'a}ndez} {et~al.}(2010){Fuentes-Fern{\'a}ndez},
  {Parnell}, \& {Hood}}]{Fuentes10}
{Fuentes-Fern{\'a}ndez}, J., {Parnell}, C.~E., \& {Hood}, A.~W. 2010, \aap,
  514, A90

\bibitem[{{Fuentes-Fern{\'a}ndez} {et~al.}(2011){Fuentes-Fern{\'a}ndez},
  {Parnell}, \& {Hood}}]{Fuentes11}
{Fuentes-Fern{\'a}ndez}, J., {Parnell}, C.~E., \& {Hood}, A.~W. 2011, \aap,
  536, A32

\bibitem[{{Fuentes-Fern{\'a}ndez}
  {et~al.}(2012{\natexlab{a}}){Fuentes-Fern{\'a}ndez}, {Parnell}, {Hood},
  {Priest}, \& {Longcope}}]{Fuentes12a}
{Fuentes-Fern{\'a}ndez}, J., {Parnell}, C.~E., {Hood}, A.~W., {Priest}, E.~R.,
  \& {Longcope}, D.~W. 2012{\natexlab{a}}, Physics of Plasmas, 19, 022901

\bibitem[{{Fuentes-Fern{\'a}ndez}
  {et~al.}(2012{\natexlab{b}}){Fuentes-Fern{\'a}ndez}, {Parnell}, \&
  {Priest}}]{Fuentes12b}
{Fuentes-Fern{\'a}ndez}, J., {Parnell}, C.~E., \& {Priest}, E.~R.
  2012{\natexlab{b}}, Physics of Plasmas, 19, 072901

\bibitem[{{Fukao} \& {Tsuda}(1973)}]{Fukao73}
{Fukao}, S. \& {Tsuda}, T. 1973, \planss, 21, 1151

\bibitem[{{Haynes} {et~al.}(2007){Haynes}, {Parnell}, {Galsgaard}, \&
  {Priest}}]{Haynes07}
{Haynes}, A.~L., {Parnell}, C.~E., {Galsgaard}, K., \& {Priest}, E.~R. 2007,
  Royal Society of London Proceedings Series A, 463, 1097

\bibitem[{{Hesse} \& {Schindler}(1988)}]{Hesse88}
{Hesse}, M. \& {Schindler}, K. 1988, Journal of Geophysical Research, 93, 5559

\bibitem[{{Hornig} \& {Priest}(2003)}]{Hornig03}
{Hornig}, G. \& {Priest}, E. 2003, Physics of Plasmas, 10, 2712

\bibitem[{{Klapper}(1997)}]{Klapper97}
{Klapper}, I. 1997, Physics of Plasmas, 5, 910

\bibitem[{{Longcope}(2001)}]{Longcope01}
{Longcope}, D.~W. 2001, Physics of Plasmas, 8, 5277

\bibitem[{{Longcope} \& {Cowley}(1996)}]{Longcope96}
{Longcope}, D.~W. \& {Cowley}, S.~C. 1996, Physics of Plasmas, 3, 2885

\bibitem[{{Longcope} \& {Parnell}(2009)}]{Longcope09}
{Longcope}, D.~W. \& {Parnell}, C.~E. 2009, \solphys, 254, 51

\bibitem[{{Masson} {et~al.}(2012){Masson}, {Aulanier}, {Pariat}, \&
  {Klein}}]{Masson12}
{Masson}, S., {Aulanier}, G., {Pariat}, E., \& {Klein}, K.-L. 2012, \solphys,
  276, 199

\bibitem[{{Masson} {et~al.}(2009){Masson}, {Pariat}, {Aulanier}, \&
  {Schrijver}}]{Masson09}
{Masson}, S., {Pariat}, E., {Aulanier}, G., \& {Schrijver}, C.~J. 2009, \apj,
  700, 559

\bibitem[{{Pariat} {et~al.}(2009){Pariat}, {Antiochos}, \& {DeVore}}]{Pariat09}
{Pariat}, E., {Antiochos}, S.~K., \& {DeVore}, C.~R. 2009, in AAS/Solar Physics
  Division Meeting, Vol.~40, AAS/Solar Physics Division Meeting, 32.01

\bibitem[{{Pariat} {et~al.}(2010){Pariat}, {Antiochos}, \& {DeVore}}]{Pariat10}
{Pariat}, E., {Antiochos}, S.~K., \& {DeVore}, C.~R. 2010, \apj, 714, 1762

\bibitem[{{Parker}(1957)}]{Parker57}
{Parker}, E.~N. 1957, \jgr, 62, 509

\bibitem[{{Parnell} {et~al.}(2008){Parnell}, {Haynes}, \&
  {Galsgaard}}]{Parnell08}
{Parnell}, C.~E., {Haynes}, A.~L., \& {Galsgaard}, K. 2008, \apj, 675, 1656

\bibitem[{{Parnell} {et~al.}(2010{\natexlab{a}}){Parnell}, {Haynes}, \&
  {Galsgaard}}]{Parnell10a}
{Parnell}, C.~E., {Haynes}, A.~L., \& {Galsgaard}, K. 2010{\natexlab{a}},
  Journal of Geophysical Research (Space Physics), 115, 2102

\bibitem[{{Parnell} {et~al.}(2010{\natexlab{b}}){Parnell}, {Maclean}, \&
  {Haynes}}]{Parnell10b}
{Parnell}, C.~E., {Maclean}, R.~C., \& {Haynes}, A.~L. 2010{\natexlab{b}},
  \apjl, 725, L214

\bibitem[{{Parnell} {et~al.}(1997){Parnell}, {Neukirch}, {Smith}, \&
  {Priest}}]{Parnell97}
{Parnell}, C.~E., {Neukirch}, T., {Smith}, J.~M., \& {Priest}, E.~R. 1997,
  Geophysical and Astrophysical Fluid Dynamics, 84, 245

\bibitem[{{Parnell} {et~al.}(1996){Parnell}, {Smith}, {Neukirch}, \&
  {Priest}}]{Parnell96}
{Parnell}, C.~E., {Smith}, J.~M., {Neukirch}, T., \& {Priest}, E.~R. 1996,
  Physics of Plasmas, 3, 759

\bibitem[{{Pontin} {et~al.}(2011){Pontin}, {Al-Hachami}, \&
  {Galsgaard}}]{Pontin11}
{Pontin}, D.~I., {Al-Hachami}, A.~K., \& {Galsgaard}, K. 2011, \aap, 533, A78

\bibitem[{{Pontin} {et~al.}(2007{\natexlab{a}}){Pontin}, {Bhattacharjee}, \&
  {Galsgaard}}]{Pontin07a}
{Pontin}, D.~I., {Bhattacharjee}, A., \& {Galsgaard}, K. 2007{\natexlab{a}},
  Physics of Plasmas, 14, 052106

\bibitem[{{Pontin} {et~al.}(2007{\natexlab{b}}){Pontin}, {Bhattacharjee}, \&
  {Galsgaard}}]{Pontin07b}
{Pontin}, D.~I., {Bhattacharjee}, A., \& {Galsgaard}, K. 2007{\natexlab{b}},
  Physics of Plasmas, 14, 052109

\bibitem[{{Pontin} \& {Craig}(2005)}]{Pontin05b}
{Pontin}, D.~I. \& {Craig}, I.~J.~D. 2005, Physics of Plasmas, 12, 072112

\bibitem[{{Pontin} \& {Galsgaard}(2007)}]{Pontin07c}
{Pontin}, D.~I. \& {Galsgaard}, K. 2007, Journal of Geophysical Research (Space
  Physics), 112, 3103

\bibitem[{{Pontin} {et~al.}(2004){Pontin}, {Hornig}, \& {Priest}}]{Pontin04}
{Pontin}, D.~I., {Hornig}, G., \& {Priest}, E.~R. 2004, Geophysical and
  Astrophysical Fluid Dynamics, 98, 407

\bibitem[{{Pontin} {et~al.}(2005){Pontin}, {Hornig}, \& {Priest}}]{Pontin05a}
{Pontin}, D.~I., {Hornig}, G., \& {Priest}, E.~R. 2005, Geophysical and
  Astrophysical Fluid Dynamics, 99, 77

\bibitem[{{Priest} \& {D{\'e}moulin}(1995)}]{Priest95}
{Priest}, E.~R. \& {D{\'e}moulin}, P. 1995, \jgr, 100, 23443

\bibitem[{{Priest} {et~al.}(2003){Priest}, {Hornig}, \& {Pontin}}]{Priest03}
{Priest}, E.~R., {Hornig}, G., \& {Pontin}, D.~I. 2003, Journal of Geophysical
  Research, 108, 1285

\bibitem[{{Priest} \& {Pontin}(2009)}]{Priest09}
{Priest}, E.~R. \& {Pontin}, D.~I. 2009, Physics of Plasmas, 16, 122101

\bibitem[{{Restante} {et~al.}(2009){Restante}, {Aulanier}, \&
  {Parnell}}]{Restante09}
{Restante}, A.~L., {Aulanier}, G., \& {Parnell}, C.~E. 2009, \aap, 508, 433

\bibitem[{{Rickard} \& {Titov}(1996)}]{Rickard96}
{Rickard}, G.~J. \& {Titov}, V.~S. 1996, \apj, 472, 840

\bibitem[{{Schindler} {et~al.}(1988){Schindler}, {Hesse}, \&
  {Birn}}]{Schindler88}
{Schindler}, K., {Hesse}, M., \& {Birn}, J. 1988, \jgr, 93, 5547

\bibitem[{{Sweet}(1958)}]{Sweet58}
{Sweet}, P.~A. 1958, in IAU Symposium, Vol.~6, Electromagnetic Phenomena in
  Cosmical Physics, ed. {B.~Lehnert}, 123

\bibitem[{{Wilmot-Smith} {et~al.}(2009){Wilmot-Smith}, {Hornig}, \&
  {Pontin}}]{Wilmot09}
{Wilmot-Smith}, A.~L., {Hornig}, G., \& {Pontin}, D.~I. 2009, \apj, 704, 1288

\bibitem[{{Wyper} \& {Jain}(2010)}]{Wyper10}
{Wyper}, P. \& {Jain}, R. 2010, Physics of Plasmas, 17, 092902

\end{thebibliography}

\end{document}